\documentclass{elsart1p}
\usepackage{graphicx}
\usepackage{amssymb}

\newcommand{\lm}{\Lambda}
\newcommand{\lb}{\Lambda_{\rm b}}
\newcommand{\be}{\begin{equation}}
\newcommand{\ee}{\end{equation}}
\newcommand{\vnn}{V_{\rm NN}}
\newcommand{\vlowk}{V_{{\rm low}\,k}}

\newcommand{\fmi}{\, {\rm fm}^{-1}}

\begin{document}

\begin{frontmatter}
\title{Nuclei and chiral dynamics}

\author{Achim Schwenk}
\address{TRIUMF, 4004 Wesbrook Mall, Vancouver, BC, V6T 2A3, Canada}
\ead{schwenk@triumf.ca}

\begin{abstract}
Nuclear theory has entered an exciting era. This is 
due to advances on many fronts, including the development of 
effective field theory and the renormalization group for nuclear
forces, advances in ab-initio methods for nuclear structure, an
effort to develop a universal density functional based on microscopic
interactions, and the application of large-scale computing resources.
I~discuss their impact, recent highlights and the frontiers in
understanding and predicting nuclei and the structure of 
strongly-interacting matter based on chiral interactions.
\end{abstract}

\end{frontmatter}

\section{Introduction}

The physics of strong interactions spans from new structures
in neutron-rich nuclei, universal properties in dilute neutron
matter and ultracold atoms, to the extremes reached in neutron
stars and supernovae. In this talk, I discuss the developments
to understand and predict nuclei and matter at the extremes
based on chiral effective field theory (EFT) and renormalization
group (RG) interactions.

\section{Effective field theory and the renormalization group
for nuclear forces}

\begin{figure}[t]
\begin{center}
\includegraphics[scale=0.325,clip=]{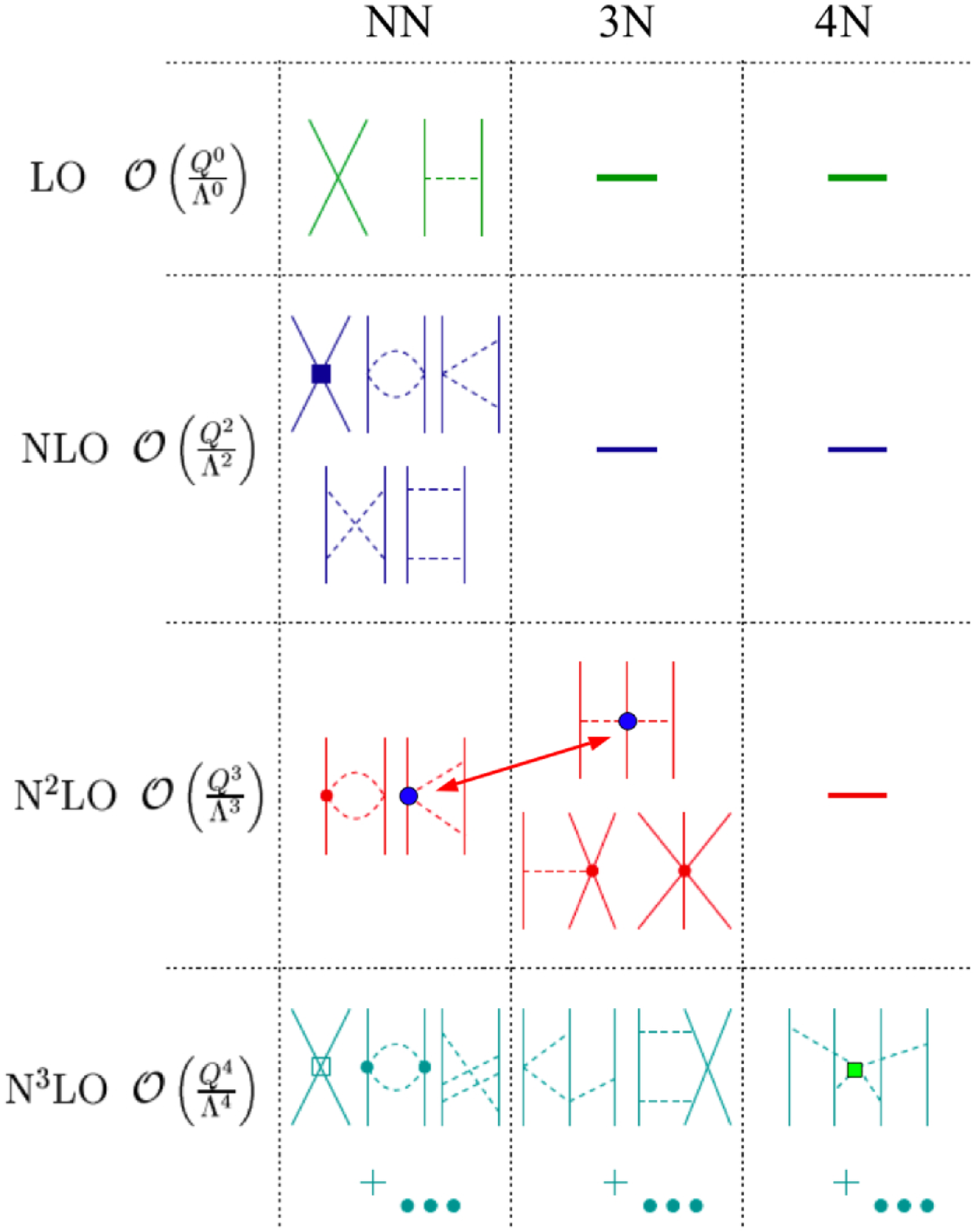}
\hspace*{6pt}
\raisebox{-6pt}{%
\includegraphics[scale=0.625,clip=]{S_P_only_n3lo.eps}}
\end{center}
\caption{Chiral EFT for nuclear forces (left) and neutron-proton
phase shifts in S- and P-waves (right) at N$^3$LO (shaded bands 
and dashed lines) in comparison to NN scattering (points). For
details see Ref.~\cite{RMP}.\label{chiralEFT}}
\vspace*{-2mm}
\end{figure}

The forces between nucleons depend on a resolution scale, which 
we denote by a generic momentum cutoff $\lm$, and are given by 
an effective theory for scale-dependent two-nucleon $\vnn(\lm)$
and corresponding many-nucleon interactions $V_{\rm 3N}(\lm),
V_{\rm 4N}(\lm), \ldots$~\cite{pionless,chiral,Vlowk,RMP}. This
scale dependence is analogous to the scale dependence of parton 
distribution functions.
At very low momenta $Q \ll m_\pi$, the 
details of pion exchanges are not resolved and nuclear forces can
be systematically expanded in contact interactions and their 
derivatives~\cite{pionless}. The corresponding pionless EFT is 
very successful for capturing universal large scattering-length
physics (with improvements by including effective range and 
higher-order terms) in dilute neutron matter and reactions
at astrophysical energies~\cite{pionless,BH,Dick,dEFT}.

For most nuclei, the typical momenta are
$Q \sim m_\pi$ and therefore pion exchanges are included explicitly
in nuclear forces. In chiral EFT~\cite{pionless,chiral,RMP}, nuclear 
interactions are organized in a systematic expansion in powers of
$Q/\lb$, where $\lb$ denotes the breakdown scale, roughly $\lb \sim
m_\rho$. As shown in Fig.~\ref{chiralEFT}, at a given order this 
includes contributions from one- or multi-pion exchanges and from
contact interactions, with short-range couplings that depend on 
the resolution scale $\lm$ and for each $\lm$ are fit to data 
(experiment captures all short-range effects).
The chiral expansion explains the phenomenological hierarchy of 
many-body forces and provides a consistent theory for multi-pion
and pion-nucleon systems, as well as for electroweak operators
(see talk by D.~Gazit). As a result, there are only two new
couplings for 3N and 4N interactions up to N$^3$LO. However, some
open questions remain: understanding the power counting with 
singular pion exchanges~\cite{counting}, including Delta degrees
of freedom, and the counting of $1/m_{\rm N}$ corrections.

Chiral EFT enables a direct connection to the underlying theory
through full lattice QCD simulations,
see for example Ref.~\cite{lattice}. This can constrain long-range
pion-nucleon couplings~\cite{Edwards}, the pion-mass dependence of
nuclear forces~\cite{Beane}, and has the potential to access
experimentally difficult observables, such as three-neutron 
properties. In addition, there are first quenched lattice QCD 
results for NN potentials~\cite{Hatsuda} in a quasi-local scheme.

\begin{figure}[t]
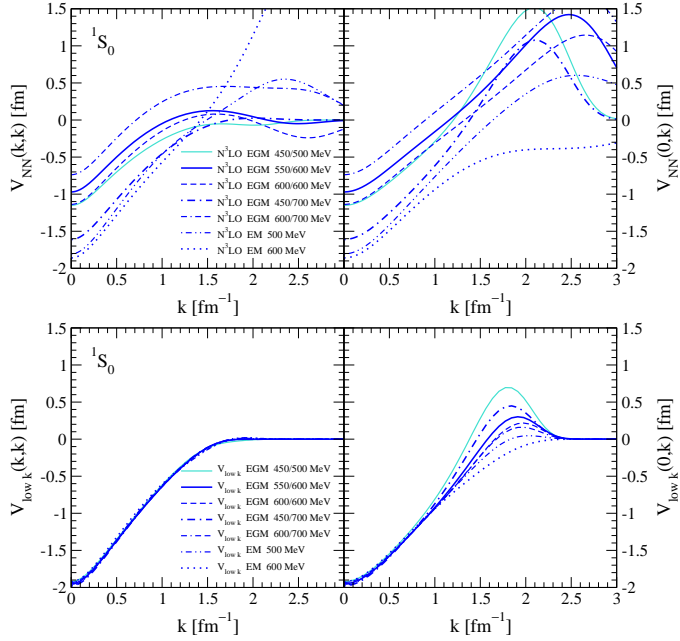

\begin{center}
\includegraphics[scale=0.32,clip=]{vnn_chiral_1s0.eps} \\
\includegraphics[scale=0.32,clip=]{vlowk_chiral_1s0.eps}
\end{center}
\vspace*{-1.5mm}
\caption{Diagonal (left) and off-diagonal (right) momentum-space 
matrix elements of different chiral EFT interactions at 
N$^3$LO~\cite{EM,EGM} (upper figures) and after RG evolution to
low-momentum interactions~$\vlowk$~\cite{Vlowk,smooth}
(lower figures) for a smooth
regulator with $\Lambda = 2.0 \fmi$.\label{universality}}
\vspace*{-0.5mm}
\end{figure}

In Fig.~\ref{universality}, we show the different chiral EFT
interactions at N$^3$LO of Entem and Machleidt (EM)~\cite{EM} 
and of Epelbaum et al. (EGM)~\cite{EGM}. These accurately
reproduce low-energy NN scattering, see Fig.~\ref{chiralEFT}.
Using the RG~\cite{Vlowk,smooth}, we can change the resolution 
scale in chiral EFT interactions and evolve N$^3$LO potentials
to low-momentum interactions $\vlowk$ with lower cutoffs. The 
RG preserves long-range pion exchanges and includes subleading
contact interactions, so that NN scattering observables and
deuteron properties are reproduced~\cite{smooth}. As shown in
Fig.~\ref{universality}, the resulting low-momentum interactions
become universal and the RG evolution weakens the off-diagonal
coupling between low and high momenta. This decoupling can also
be achieved using similarity RG transformations towards 
band-diagonal~\cite{SRG1} or block-diagonal~\cite{SRG2} interactions 
in momentum space.

Changing the cutoff leaves observables unchanged by construction, but
shifts contributions between the interaction strengths and the sums 
over intermediate states in loop integrals. The evolution of chiral 
EFT interactions to lower resolution is beneficial, because these
shifts can weaken or largely eliminate sources of nonperturbative
behavior such as strong short-range repulsion and short-range tensor 
forces~\cite{Born}. Lower cutoffs need smaller bases in many-body 
calculations, leading to improved convergence for nuclei~\cite{NCSM}.

Chiral EFT interactions become more accurate with higher orders
and the RG cutoff variation estimates theoretical uncertainties
due to higher-order contributions (as shown for phase shifts in
Fig.~\ref{chiralEFT}), due to neglected many-body interactions
or an incomplete many-body treatment. This is a powerful tool
for extrapolations to the extremes and for assessing the
uncertainties of pivotal matrix elements needed in fundamental 
symmetry tests (for example, for 
isospin-symmetry-breaking corrections to superallowed decays,
for double-beta decay, or for octupole enhancement factors 
of electric dipole moments).

\section{Nuclear forces and the structure of matter}

\begin{figure}[t]
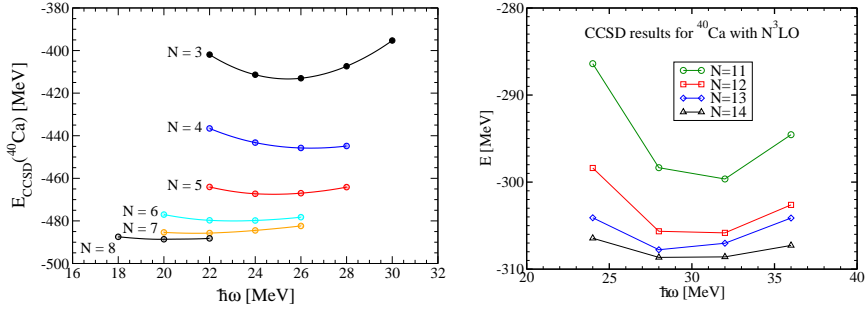

\begin{center}
\includegraphics[scale=0.225,clip=]{ca40_hwdep.eps}
\hspace*{6pt}
\raisebox{2pt}{%
\includegraphics[scale=0.185,clip=]{ca40_n3lo500_NL.eps}}
\end{center}
\vspace*{-1.5mm}
\caption{CCSD ground-state energy of $^{40}$Ca as function of the
oscillator frequency with increasing basis size~$N$,
based on an RG-evolved low-momentum NN interaction
(left)~\cite{VlowkCC} and a chiral N$^3$LO NN potential 
(right)~\cite{CCchiral}. The difference highlights the importance
of 3N interactions for ground-state energies.\label{CCfig}}
\end{figure}

Advancing ab-initio methods to heavier and neutron-rich nuclei
and to reactions are frontiers in nuclear theory. Coupled-cluster
(CC) theory is the prime method for systems with up to 100 
electrons in quantum chemistry~\cite{Bar07} and a powerful method
for nuclei for which a closed-shell reference state provides a good
starting point~\cite{Gour06,Hag06}. Combined with rapid convergence
for low-momentum interactions, CC theory has pushed the limits of
accurate calculations to medium-mass nuclei and set new benchmarks 
for $^{16}$O and $^{40}$Ca~\cite{VlowkCC}. Using an angular-momentum-%
coupled scheme, it is possible to extend CC theory
to gigantic spaces (15 major shells on a single processor) and to
obtain near-converged ground-state energies for spherical nuclei, 
$^{40}$Ca, $^{48}$Ca, and $^{48}$Ni, based on a chiral N$^3$LO NN
potential~\cite{CCchiral}. The CC developments for medium-mass
nuclei are shown in Fig.~\ref{CCfig}. Future milestones can
be set by first ab-initio calculations with 3N interactions for
medium-mass nuclei, for neutron-rich tin isotopes, and for the
neutron radius of $^{208}$Pb (for the JLab/PREX experiment) based
on chiral EFT.

\begin{figure}[t]
\begin{center}
\includegraphics[scale=0.32,clip=]{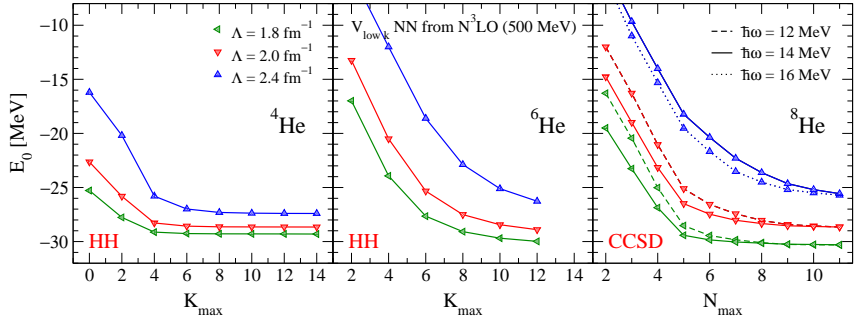}
\end{center}
\vspace*{-1.5mm}
\caption{Convergence as a function of the basis size for the
HH $^{4}$He and $^{6}$He ground-state
energies and for the CCSD ground-state energy of $^8$He based
on chiral low-momentum NN interactions $\vlowk$ for a range of
cutoffs $\Lambda = 1.8$, $2.0$ and $2.4 \fmi$. For details see
Ref.~\cite{HHCC}.\label{4-8He}}
\end{figure}

\begin{figure}[t]
\begin{center}
\includegraphics[scale=0.45,clip=]{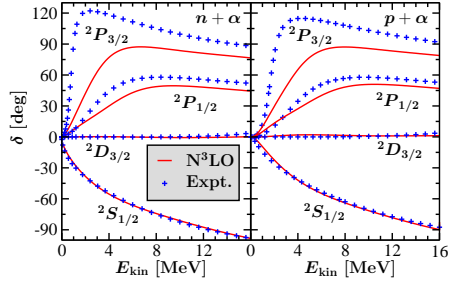}
\end{center}
\vspace*{-1.5mm}
\caption{Nucleon-$^4$He phase shifts based on a chiral N$^3$LO 
NN potential using the resonating-group method combined 
with the NCSM, in comparison to an $R$-matrix analysis of data.
For details see Ref.~\cite{Sofia}.\label{N4He}}
\vspace*{-1.5mm}
\end{figure}

Recently, a combination of nuclear and atomic physics techniques
led to the first precision measurements of masses and charge radii
of the helium halo nuclei, $^6$He~\cite{6He} and $^8$He~\cite{8He},
with two or four weakly-bound neutrons forming an extended halo
around the $^4$He core. In Fig.~\ref{4-8He}, we show results for
the ground-state energies of helium nuclei based on chiral
low-momentum NN interactions~\cite{HHCC}. This combines the RG
evolution with the exact hyperspherical-harmonics (HH) expansion for
$^6$He and CC theory for $^8$He (see also Ref.~\cite{Hag06}),
which have the correct asymptotic behavior of the wave function.
The cutoff variation in Fig.~\ref{4-8He} highlights the importance
of 3N interactions. For all studied cutoffs, the NN-only results 
underbind $^8$He~\cite{Hag06,HHCC}. The helium isotopes therefore
probe 3N effects beyond the repulsion in infinite nuclear and 
neutron matter (see Figs.~\ref{nucmatt} and~\ref{nm}), and we 
expect the necessary attractive contribution to $^8$He is due
to spin-orbit effects (which largely average out in infinite
matter). The importance of 3N forces for spin-orbit physics
is also apparent in the nucleon-$^4$He P-wave-splitting in 
Fig.~\ref{N4He}. These are the first ab-initio calculations of
(elastic) nuclear reactions based on chiral EFT interactions~\cite{Sofia}.

\begin{figure}[t]
\begin{center}
\includegraphics[scale=0.22,clip=]{tjon-line.eps}
\hspace*{6pt}
\raisebox{-5pt}{%
\includegraphics[scale=0.3,clip=]{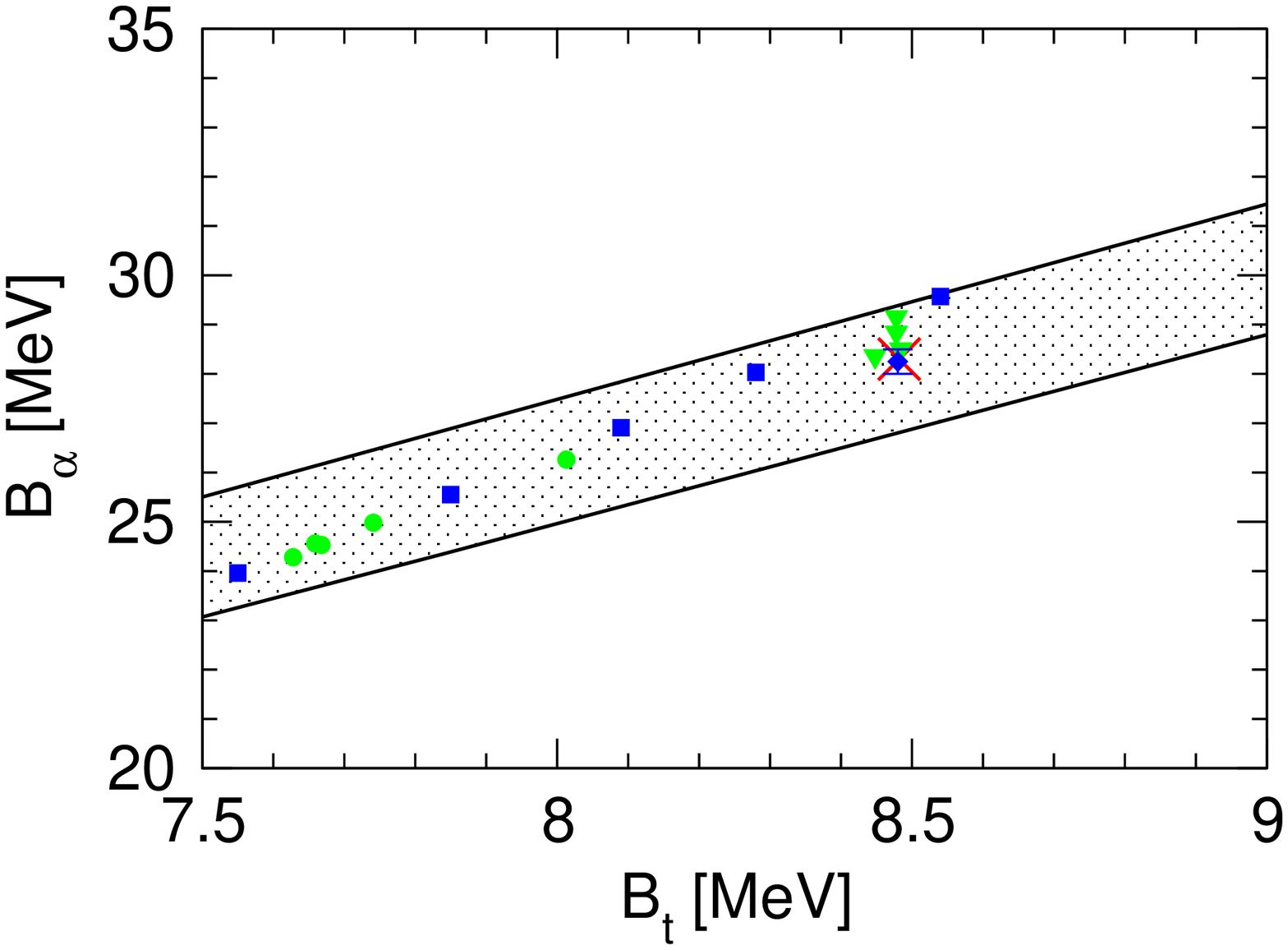}}
\end{center}
\vspace*{-1.5mm}
\caption{Correlation of $^3$H and $^4$He binding energies.
The cutoff dependence obtained from low-momentum NN interactions 
$\vlowk$ (left figure) explains the empirical (solid) Tjon
line~\cite{Vlowk3N}. This correlation is driven by large scattering
lengths, as demonstrated by the band (right figure) obtained in 
pionless EFT~\cite{Hans}.\label{Tjon}}
\end{figure}

\section{Three-nucleon interactions: a frontier}

Three-nucleon forces are crucial for binding energies and
radii, play a central role for spin-orbit effects, for the
evolution of nuclear structure with isospin, and they drive
the density dependence of nucleonic matter~\cite{Tokyo}.
In addition, 3N forces are required for
renormalization. As shown in Fig.~\ref{Tjon}, neglecting 
3N interactions leads to a universal correlation
(empirically known as Tjon-line) between the $^3$H and $^4$He
binding energies~\cite{Vlowk3N,Hans}.

\begin{figure}[t]
\begin{center}
\includegraphics[scale=0.21,clip=]{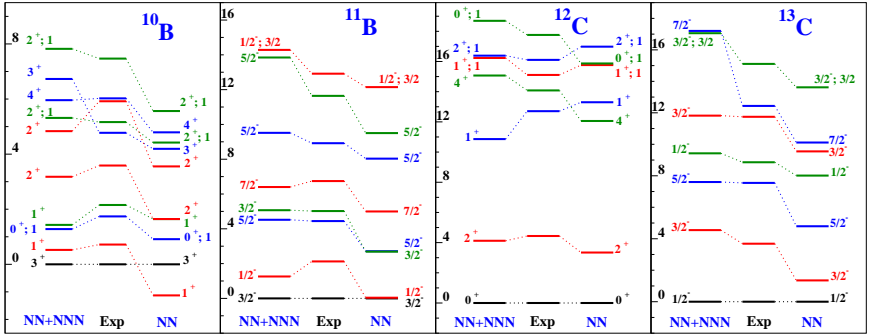}
\end{center}
\vspace*{-1.5mm}
\caption{Excitation energies in MeV of light nuclei obtained
in the No-Core Shell Model (NCSM) with chiral EFT interactions
(NN to N$^3$LO and 3N to N$^2$LO)~\cite{NCSMchiral}.\label{pshell}}
\end{figure}

In chiral EFT without explicit Deltas, 3N interactions start
at N$^2$LO~\cite{chiral3N} (see Fig.~\ref{chiralEFT}) and 
their contributions are given diagrammatically by
\vspace*{-2mm}
\be
\nonumber
\hspace*{35mm}
\parbox[c]{0pt}{%
\includegraphics[scale=0.21,clip=]{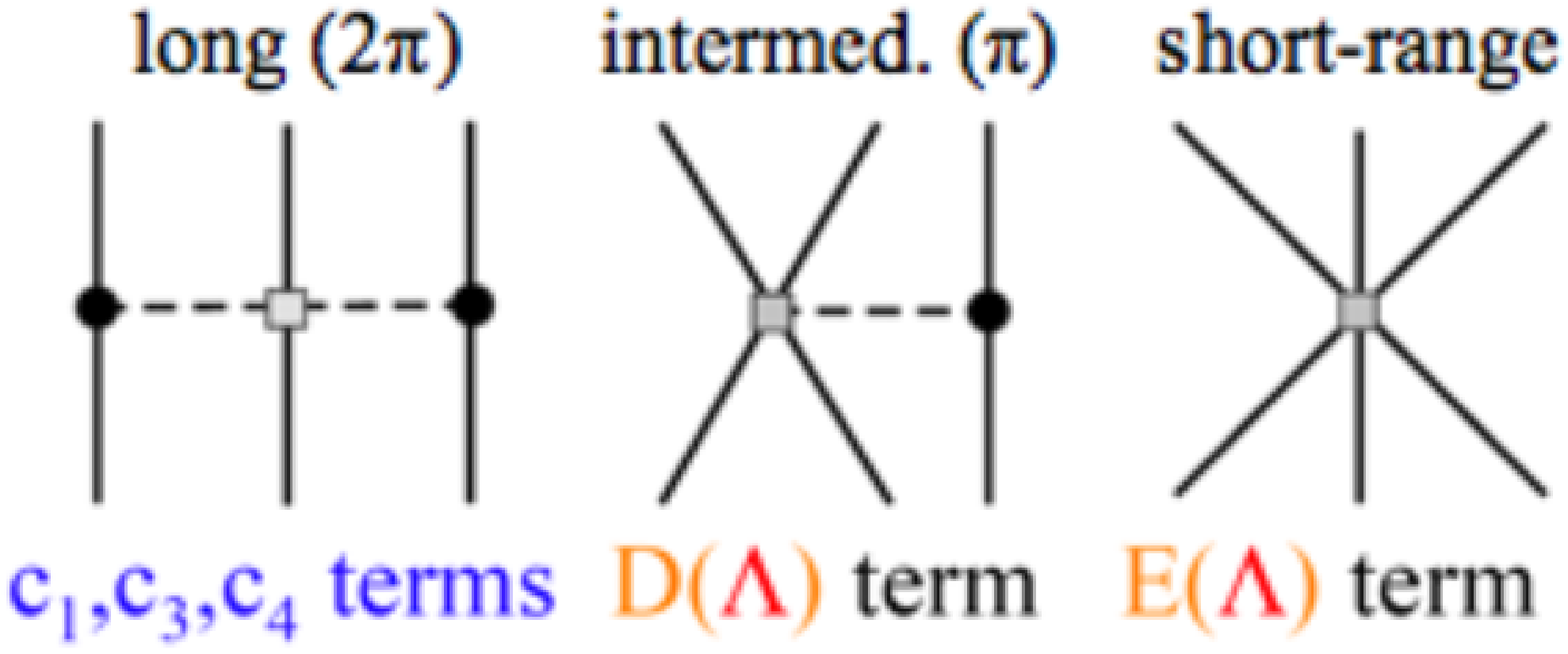}}
\ee
The long-range two-pion-exchange part is determined by the 
couplings $c_1, c_3, c_4$, which have been constrained in 
the $\pi$N or NN system, and the remaining D-~and E-term
couplings are usually fitted to the $^3$H binding energy 
and another observable in $A \geqslant 3$. In particular,
$c_3$ and $c_4$ are important for nuclear structure and have
large uncertainties (see Fig.~\ref{nm} for the impact on
neutron matter). The leading 3N interactions generally
improve the agreement of theory with experiment in
nucleon-deuteron scattering (see talk by H.~Wita{\l}a)
and in spectra of light nuclei~\cite{NCSMchiral} shown
in Fig.~\ref{pshell}. The subleading 3N interactions
at N$^3$LO are parameter free and several parts have been
calculated~\cite{sub3N} (see talk by E.~Epelbaum). They 
decrease the $c_i$ values, include two-pion--one-pion-exchange
and pion-ring diagrams, spin-orbit forces and contributions
that involve NN contacts.

\begin{figure}[t]
\begin{center}
\includegraphics[scale=0.22,clip=]{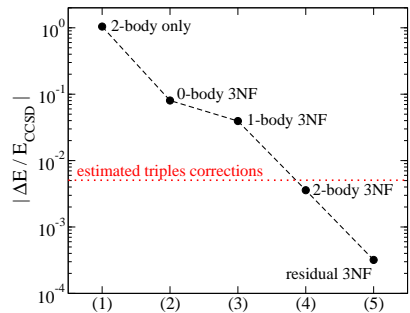}
\end{center}
\vspace*{-1.5mm}
\caption{Relative contributions $|\Delta E / E|$ to the binding energy 
of $^4$He at the CCSD level from $\vlowk$ as well as normal-ordered
0-, 1-, 2-body and residual 3-body parts of the 3N 
interaction~\cite{CC3N}.\label{3Nparts}}
\end{figure}

Since chiral EFT is a complete low-momentum basis, we have 
constructed 3N interactions $V_{\rm 3N}(\lm)$, corresponding to 
RG-evolved interactions, by fitting the leading D-~and E-term
couplings to the $^3$H binding energy and the $^4$He binding 
energy~\cite{Vlowk3N} or radius~\cite{new3N} for a range of
cutoffs. For lower cutoffs, low-momentum 3N interactions 
become perturbative in light nuclei~\cite{Vlowk3N}, and
their expectation values are consistent with power counting
estimates. Future work includes evolving chiral 3N interactions.
Using the similarity RG, the evolution of three-body forces is
feasible with very encouraging results in 1d~\cite{1d}.

In Fig.~\ref{3Nparts}, first CC results with 3N forces show that 
low-momentum 3N interactions are accurately treated as effective
0-, 1- and 2-body terms, and that residual 3N forces can be
neglected~\cite{CC3N}. This is very promising for developing
tractable approximations to handle many-body interactions and
supports the idea that phenomenological adjustments in shell 
model interactions (``monopole shifts'') are due to 3N 
contributions~\cite{Zuker}. These monopole shifts are enhanced in 
neutron-rich nuclei and have a pivotal impact on shell closures 
and the location of the neutron drip line. For example, shell 
model interactions obtained from microscopic NN (including
chiral N$^3$LO) potentials miss the location of the neutron drip
line in oxygen ($^{28}$O compared to the correct $^{24}$O). First 
results indicate that the leading chiral 3N contributions,
dominated by the $c_i$ terms (Delta-hole effects),
can provide the necessary repulsive shifts~\cite{mono}.

\section{Advances for nuclear matter and towards
a universal density functional}

\begin{figure}[t]
\begin{center}
\includegraphics[scale=0.3,clip=]{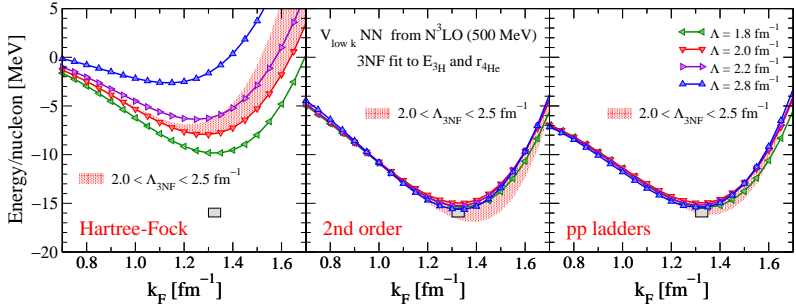}
\end{center}
\vspace*{-1.5mm}
\caption{Nuclear matter energy per nucleon versus Fermi momentum
$k_{\rm F}$ at the Hartree-Fock level (left) and including second-order
(middle) and particle-particle ladder contributions (right)~\cite{new3N},
based on evolved N$^3$LO NN potentials and 3N fits to $E_{^3{\rm H}}$ and 
$r_{^4{\rm He}}$. Theoretical uncertainties are estimated
by the NN (lines) and 3N (band) cutoff variation; the empirical 
saturation point is given by the box.\label{nucmatt}}
\end{figure}

\begin{figure}[t]
\begin{center}
\includegraphics[scale=0.26,clip=]{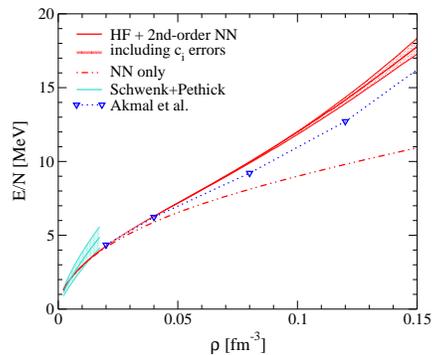}
\end{center}
\vspace*{-1.5mm}
\caption{Neutron matter energy per particle versus density based
on low-momentum NN and 3N interactions including Hartree-Fock plus
dominant second-order contributions. For details see Ref.~\cite{nm}.
\label{nm}}
\end{figure}

Low-momentum interactions offer the possibility of perturbative 
nuclear matter~\cite{nucmatt}, whereas for hard potentials, the 
coupling to high momenta renders the Bethe-Brueckner-Goldstone
expansion necessarily nonperturbative. New results~\cite{new3N}
for nuclear matter starting from chiral EFT interactions combined
with RG evolution are shown in Fig.~\ref{nucmatt}. We
have found that 3N forces drive saturation~\cite{nucmatt}
and observe a weak cutoff dependence in Fig.~\ref{nucmatt},
when second-order contributions are included. The cutoff 
dependence is improved significantly by constraining
3N interactions with the $^4$He radius (compared to Fig.~6 in 
Ref.~\cite{nucmatt}), and it is exciting that the empirical 
saturation point is predicted within theoretical 
uncertainties~\cite{new3N}.

In Fig.~\ref{nm}, we show the impact of 3N interactions on neutron
matter~\cite{nm}. At present, the uncertainties in the $c_i$ values
(mainly $c_3$ for neutron matter) overwhelm the errors due to cutoff 
variation (the band in Fig.~\ref{nm}). This also establishes a
correlation of low-momentum 3N interactions with the symmetry energy.
In addition to these advances, there are nuclear lattice simulations
for dilute neutron matter in lattice chiral EFT~\cite{nuclatt}.

\begin{figure}[t]
\begin{center}
\includegraphics[scale=0.42,clip=]{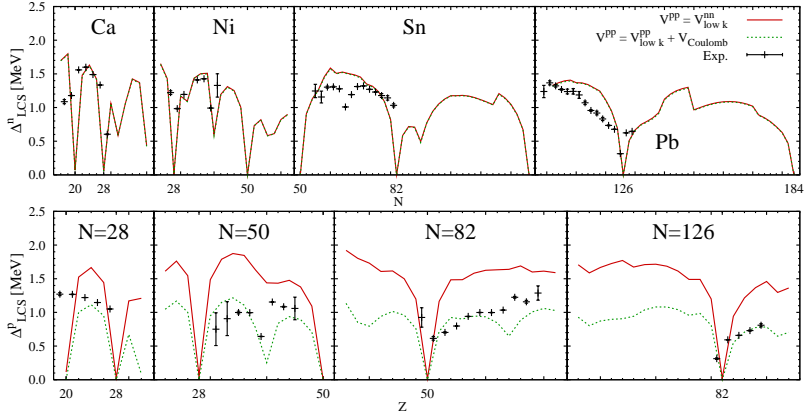}
\end{center}
\vspace*{-1.5mm}
\caption{Neutron/proton gaps (upper/lower panel) along isotopic/isotonic
chains obtained from $\vlowk$ (plus Coulomb potential for proton gaps)
as pairing interaction in density functional calculations~\cite{pairing}.
\label{gaps}}
\end{figure}

The nuclear matter results of Fig.~\ref{nucmatt} and 
Ref.~\cite{nucmatt} imply that exchange
correlations are tractable, and this motivates the derivation of
an ab-initio density functional (in a path integral formulation,
the density functional is the effective action for the density)
based on low-momentum interactions, for example, using a density 
matrix expansion~\cite{DME} (see also Ref.~\cite{Dick,Wolfram} and
the talk by W.~Weise). This is one of the goals of the SciDAC
universal nuclear energy density functional project (unedf.org).
Future work can use EFT and RG to identify new terms in the 
functional, to quantify theoretical errors for extrapolations,
and to benchmark with ab-initio methods (in particular with CC theory
for medium-mass nuclei). Results for pairing gaps from the first
non-empirical pairing functional based on $\vlowk$ are very 
promising~\cite{pairing}, as shown in Fig.~\ref{gaps}.

\begin{figure}[t]
\begin{center}
\includegraphics[scale=0.26,clip=]{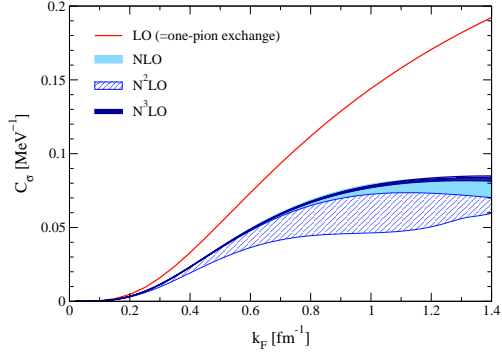}
\end{center}
\vspace*{-1.5mm}
\caption{Spin relaxation rate given by $C_\sigma$ in neutron matter
as a function of Fermi momentum $k_{\rm F}$ obtained from chiral 
EFT interactions of successively higher orders~\cite{brems}.\label{orders}}
\end{figure}

\section{Neutrino interactions and nuclear matter in astrophysics}

Neutrinos play a crucial role for the physics of stellar 
collapse, supernova explosions and neutron stars~\cite{neutrinos}
(see also the talk by M.~Liebend\"orfer). In this last Section,
we discuss first chiral EFT results for neutrino rates involving
two nucleons, such as neutrino-pair bremsstrahlung and absorption,
$N N \leftrightarrow N N \nu \overline{\nu}$, which are key for
muon and tau neutrino production and their spectra.
These processes are determined by the spin
relaxation rate $1/\tau_\sigma = C_\sigma \, (T^2 + [\omega/2\pi]^2)$
using a unified approach to neutrino interactions in nucleon 
matter~\cite{LPS,brems}. The bremsstrahlung rates are proportional to
$C_\sigma$ to good approximation.

In supernova simulations, the standard rates for bremsstrahlung
are based on the one-pion exchange (OPE) approximation to NN
interactions~\cite{HR}. This is a reasonable starting point, 
since it represents the long-range part of nuclear forces, and
for neutron matter, it is the leading-order contribution to
the spin relaxation rate in chiral EFT. In Fig.~\ref{orders}, we 
systematically go beyond the OPE approximation and show the
spin relaxation rate from chiral EFT interactions to N$^3$LO
over a wide density range~\cite{brems}. Two-pion-exchange 
interactions and shorter-range noncentral forces at NLO reduce
the neutrino rates significantly. Future work will include
neutron-proton mixtures and charged currents, to systematically
improve the neutrino physics input for astrophysical simulations.

\section{Summary}

There is an exciting and coherent effort in nuclear theory
to understand and predict nuclear systems based on effective
field theory and renormalization group interactions, which
has the potential to provide a unified and systematic 
description of light to heavy nuclei and matter in 
astrophysics based on the same interactions and with
controlled uncertainties. A common frontier is the consistent
inclusion of three-nucleon interactions.

\section{Acknowledgements}

It is a pleasure to thank the organizers for a very stimulating
meeting and for a most memorable hospitality in Israel, and
S.~Bacca, S.~Bogner, D.~Dean, B.~Friman, R.~Furnstahl, G.~Hagen,
K.~Hally, J.~Holt, A.~Nogga, T.~Otsuka, T.~Papenbrock, C.~Pethick
and T.~Suzuki for many discussions. This work was supported in
part by the Natural Sciences and Engineering Research Council (NSERC).
TRIUMF receives federal funding via a contribution agreement through 
the National Research Council of Canada.

\end{document}